\documentclass[%
preprint,
showpacs,
bibnotes,
amsmath,
amssymb,
elsearticle,
showkeys
]{revtex4-1}

\usepackage{graphicx}% Include figure files
\usepackage{dcolumn}% Align table columns on decimal point
\usepackage{bm}% bold math
\usepackage{color}
\usepackage[mathlines]{lineno}% Enable numbering of text and display math

\usepackage{fancyhdr}
\begin{document}

\def\mean#1{\left< #1 \right>}

\preprint{Article published on Journal of Computational Physics - DOI: 10.1016/j.jcp.2016.09.023}

\title{Meaningful timescales from Monte Carlo simulations of molecular systems with hard-core interactions}% Force line breaks with \\

\author{Liborio I. Costa}
%\homepage{liborio78@gmail.com}
%\affiliation{Etzbergstrasse 19c, 8405 Winterthur, Switzerland}%
\affiliation{Winterthur, Switzerland}%
\email{liborio78@gmail.com}

%\date{March 21, 2016 - working paper}% It is always \today, today,

\begin{abstract}
A new Markov Chain Monte Carlo method for simulating the dynamics of molecular systems characterized by hard-core interactions is introduced. In contrast to traditional Kinetic Monte Carlo approaches, where the state of the system is associated with minima in the energy landscape, in the proposed method, the state of the system is associated with the set of paths traveled by the atoms and the transition probabilities for an atom to be displaced are proportional to the corresponding velocities. In this way, the number of possible state-to-state transitions is reduced to a discrete set, and a direct link between the Monte Carlo time step and true physical time is naturally established. The resulting rejection-free algorithm is validated against event-driven molecular dynamics: the equilibrium and non-equilibrium dynamics of hard disks converge to the exact results with decreasing displacement size.
\end{abstract}

% PACS, the Physics and Astronomy  Classification Scheme.
%Use showkeys class option if keyword display desired

\maketitle

%\tableofcontents

%\linenumbers

\noindent \textbf{1. Introduction}

 Monte Carlo (MC) stochastic methods are currently well established as numerical tools for unraveling the complex behavior of a large variety of systems \cite{Baumgàrthner1980,Bernard2011,Yoshida2005,Potestio2013,Milchev1996,Meimaroglou2014,Munoz2003,Zuniga2015,Patti2012,Tiana2007,Wang2014}. Most MC algorithms can be divided into two main groups: methods built on the importance sampling scheme of Metropolis et al. \cite{Metropolis1953} (all denoted here as MMC \cite{Bal2014,Bernard2012,Bernard2011,Grassberger1997,LandauBinder,Mees2012,Neyts2012,Peters2012})  and kinetic Monte Carlo (KMC) methods \cite{VoterBook, Henkelman2001,Fichthorn1991}.

In MMC, the system evolves through randomly sampled configurations. Each new configuration is accepted or rejected according to specific rules to satisfy the Boltzmann equiprobability principle. The possibility of letting the system evolve through unphysical trajectories \cite{Bernard2012, Grassberger1997} makes MMC algorithms particularly efficient for calculating equilibrium properties. In fact, they are often preferred over molecular dynamics (MD) in those cases where relaxation times diverge. As representative examples, one can mention the melting transition of hard disks \cite{Bernard2011} and the conformations of the densely crowded molecular architectures \cite{Yoshida2005} typical of dendronized polymers \cite{Costa2008,Costa2011}. The price to pay in MMC is the lack of a physically meaningful timescale. In fact, although MMC can be used to obtain information about dynamics \cite{Tiana2007,Wang2014, Baumgàrthner1980, Munoz2003, Milchev1996}, often only in terms of scaling laws, it is at the same time recognized that the analogy between the MMC dynamics and actual time evolution is only superficial \cite{Baumgàrthner1980, Munoz2003, Peters2012, LandauBinder}. In the field of molecular simulations, notable exceptions are Brownian processes \cite{Kikuchi1991, Sanz2010, Patti2012} and the bond-fluctuation model for polymer chains, which provides a good approximation of the Rouse model \cite{Baumgàrthner1980, LandauBinder}. However, in both cases, the connection to physical time requires a priori knowledge, or at least an estimation, of the diffusion coefficient, which is not always easy to predict, especially in concentrated polymeric systems \cite{Costa2010}. Additionally, these cases are the exceptions rather than the rule, and a general and definitive connection between physical time and the dynamic evolution of molecular systems determined through MMC simulations is still missing despite the many efforts in this direction \cite{Sanz2010, Bal2014,Neyts2012, Mees2012}.

The correct time evolution can be obtained, in principle, through KMC methods, which apply to systems that evolve dynamically from state to state, each state being identified with a minimum in the energy landscape \cite{VoterBook, Fichthorn1991,Middleton2004}. Voter, for example \cite{Voter1986}, in one of his seminal works, used KMC to study the diffusion of rhodium clusters on Rh(100). In KMC, atomic vibrations are neglected, and, provided that one knows all possible states corresponding to the minima of the energy basins, a rate catalog for the transitions from one state to another can be build, where the rate constants are calculated using the transition state theory \cite{VoterDoll1984, VoterDoll1985}. Once the rate catalogue is given, the trajectory of the system through such states can be obtained by means of a stochastic procedure \cite{Voter1986, Fichthorn1991, VoterBook}. On the other hand, if a system undergoes unexpected reaction pathways, the task to identify and quantify all possible states and to build the corresponding rate catalogue becomes extremely challenging. To this end, advanced KMC methods have been proposed, where the rate catalogue is built on the fly \cite{Henkelman1999, Henkelman2001}. While promising, such methods add some complexity to the algorithm, and, in practice, one can never fully ensure that all states have been identified. As a consequence, to date, KMC methods can be used when the number of states is discrete and the transition probabilities for the state-to-state transitions are known or easy to identify, but they are not suitable or of difficult application for molecular systems evolving in a continuous space characterized by an infinite and unknown number of states. The reader can consult a very clear introduction on the topic by Voter for additional details on the pros and cons of KMC \cite{VoterBook}.

Therefore, in general terms, if the interest is in studying the dynamics of a molecular system with full atomistic detail (e.g. including the atomic vibrations) and without any a priori knowledge of the system behavior, one has to revert to MD. At the same time, it remains an open question whether the same type of information could be obtained by a generally valid MC. The aim of this contribution is to reduce the gap between MC and MD by proposing a new Markov Chain MC scheme that provides a reasonably accurate description of the dynamics of molecular systems without the above mentioned limitations  of KMC methods. As a first step towards this goal, the case of systems characterized by hard-core interactions is considered here, and the corresponding algorithm is conveniently indicated as Monte Carlo Molecular Dynamics (MCMD). In what follows, I first introduce the concepts underlying MCMD and the resulting algorithm. Then, the method is validated against event-driven MD in terms of both static and dynamic properties for the case of hard disks in a box. Finally, an outlook for possible future generalizations and applications is given.

\noindent \textbf{2. Analogy between motion and reaction network}

Let us consider first the equations of motion for $N$ non-interacting hard disks. In the absence of collisions, each disk $i$ moves along a straight line parallel to its velocity $v_i$ according to

\begin{equation}\label{Equation1}
dl_i/dt = v_i
\end{equation}

where $l_i$ is the distance traveled by disk $i$. The system of $N$ equations (Eq.~\ref{Equation1}) can be written in terms of the rescaled quantities $n_i$ = $l_i$/$\delta$ and $a_i$ = $v_i$/$\delta$ where the parameter $\delta$ is the “unit of displacement”, such that $n_i$ represents the number of steps $\delta$ traveled by disk $i$:

\begin{equation}\label{Equation2}
dn_i/dt = a_i
\end{equation}

At the same time, we notice that solving Eq.~\ref{Equation2} (and therefore Eq.~\ref{Equation1}) is mathematically equivalent to solving a chemical reaction network where all reactions are of the external source type

\begin{equation}\label{Equation3}
\O \xrightarrow{a_i} n_i
\end{equation}

From this view point, $n_i$ and $a_i$ would represent the number of molecules of species $i$ and the rate parameter of the $i^{th}$ reaction generating the species $i$, respectively. By treating $n_i$ as discrete variables, the probabilistic evolution of such a system is governed by the master equation

\begin{equation}\label{Equation4}
\frac {\partial P(t,\textbf{\textit{n}})} {\partial t}= \sum_{i=1}^N a_i[P(t,n_1,...,n_i-1,...,n_N)-P(t,\textbf{\textit{n}})]
\end{equation}

where $P(t,\textbf{\textit{n}})$ is the probability that the system is in state $\textbf{\textit{n}}$ = ($n_1$,...,$n_N$) at time $t$. In turn, Eq.~\ref{Equation4} can be solved using a KMC scheme, namely the stochastic simulation algorithm (SSA) \cite{Gillespie1976}. SSA is an iterative algorithm where at each iteration, a reaction channel $\mu$ is sampled proportionally to the transition probability $a_\mu$, named propensity in the terminology of the SSA, the state of the system is updated, $n_{\mu} \leftarrow n_{\mu}$+1, and the time is advanced of the MC time step \cite{note_tau}

\begin{equation}\label{Equation5}
\tau = \frac {1} {A} = 1/ \sum_{i=1}^N a_i
\end{equation}

In other words, by restricting the infinite number of possible random displacements to the discrete set of $N$ displacements parallel to the disks velocities, one can identify the state of the system at time $t$ with the number of displacements $\delta$ traveled by the disks, $P(t,\textbf{\textit{n}})$, rather than with its energy \cite{Fichthorn1991, VoterBook, Middleton2004}, and one can construct a Markov Chain process, Eq.~\ref{Equation4}, where the state-to-state transition probabilities (propensities) at time $t$ are the $N$ scaled velocities $a_i$. More importantly, a direct link between the MC time step and true physical time is established through Eq.~\ref{Equation5}. Notably, $P(t,\textbf{\textit{n}})$ approaches the deterministic limit as $n_i$ increases \cite{Gillespie1976}, i.e., the smaller the value of the unit of displacement $\delta$, the closer the stochastic evolution of $l_i$ to the deterministic solution. Note that $a_i$ has not to be confused with the acceleration, which is zero for the case of hard disks. Rather, the symbol $a_i$ represents the probability of disk $i$ to be sampled (cf. Eq.~\ref{Equation4} and the algorithm in Section 3). This choice is in line with the usual symbology used when applying KMC algorithms to reaction networks, with $a_i$ representing the so-called propensity or firing probability.\cite{Gillespie1976}

Keeping the above simple considerations in mind, one can generalize this approach by including elastic collisions between the $N$ disks and between disks and box walls and by removing the space discretization. The resulting algorithmic steps are presented and discussed as follows.

\noindent \textbf{3. MCMD Algorithm}

1) The system is first initialized at $t$ = 0 by fixing the maximum displacement $\delta$ and assigning to each disk $i$ the initial positions ($x_i$,$y_i$), and velocities $\textbf{\textit{v}}_i=(v_{ix}, v_{iy})$, with $v_i=(v_{ix}^2+v_{iy}^2)^{1/2}$.

2) The displacement of the current MC step is calculated as $r_1\delta$, where $r_1$ is a random number drawn from the uniform distribution in the unit interval. The atom propensities are $a_i$ = $v_i/(r_1\delta)$ and $A=\sum_{i=1}^N a_i$.

3) The disk to be displaced is sampled by drawing a second random number, $r_2$, drawn from the uniform distribution in the unit interval, as the disk $\mu$ that satisfies the inequalities: $\sum_{i=1}^{\mu-1} a_i < r_2A \leq \sum_{i=1}^{\mu} a_i$.

4) The minimum distances to a wall and to a pair collision for disk $\mu$, $\delta_{wall}$ and $\delta_{pair}$, respectively, are calculated along the direction of $v_\mu$, keeping all other disks fixed (Supplemental Material \cite{Supplemental}).

5a) \textbf{If} $r_1\delta < \min(\delta_{wall}, \delta_{pair})$, there are no collisions in the current step. The positions are updated and the time step is calculated:
\begin{equation}\label{Equation6}
x_\mu \leftarrow x_\mu+r_1\delta v_{\mu x}/v_\mu;  \;\;\;\;\;   y_\mu \leftarrow y_\mu+r_1\delta v_{\mu y}/v_\mu
\end{equation}
\begin{equation}\label{Equation7}
\tau=1/A
\end{equation}

5b) \textbf{Else}, a collision takes place. In this case positions and time step are
\begin{subequations}\label{Equation8}
\begin{align}
x_\mu \leftarrow x_\mu+\min(\delta_{wall}, \delta_{pair}) v_{\mu x}/v_\mu \\
y_\mu \leftarrow y_\mu+\min(\delta_{wall}, \delta_{pair}) v_{\mu y}/v_\mu
\end{align}
\end{subequations}
\begin{equation}\label{Equation9}
\tau=\min(\delta_{wall}, \delta_{pair})/\sum_{i=1}^{N} v_i
\end{equation}\\

Following the collision event, the velocity/ies of the colliding disk/s is/are updated exactly as in the conventional event driven MD, i.e., assuming a perfectly elastic collision preserving kinetic energy and quantity of motion \cite{Haile1997}.

6) The time is advanced as $t \leftarrow t+\tau$, and the algorithm is iterated from step 2 until  $t<t_{end}$.\\

Thus, steps 2 to 6 construct a Markov Chain process, which is terminated in step 5b when a collision takes place. In 5b the time step is calculated on the line of Eq.~\ref{Equation5} as the ratio between the distance traveled and the sum of all current velocities. In short, the evolution calculated by the MCMD is built as a sequence of Markov Chains connected by collision events. In step 2, a random displacement is used rather than a fixed one. In this way, the disks move in a continuous space rather than in a discrete one. We then proceed with the validation of the MCMD. Simulations were performed with $N = 64$ disks in a box of size 1x1. The computational details are given in the Supplemental Material \cite{Supplemental}.

\noindent \textbf{4. Results}\\
\noindent \textbf{4.1 Static properties}

\begin{figure}
\includegraphics[scale=0.55]{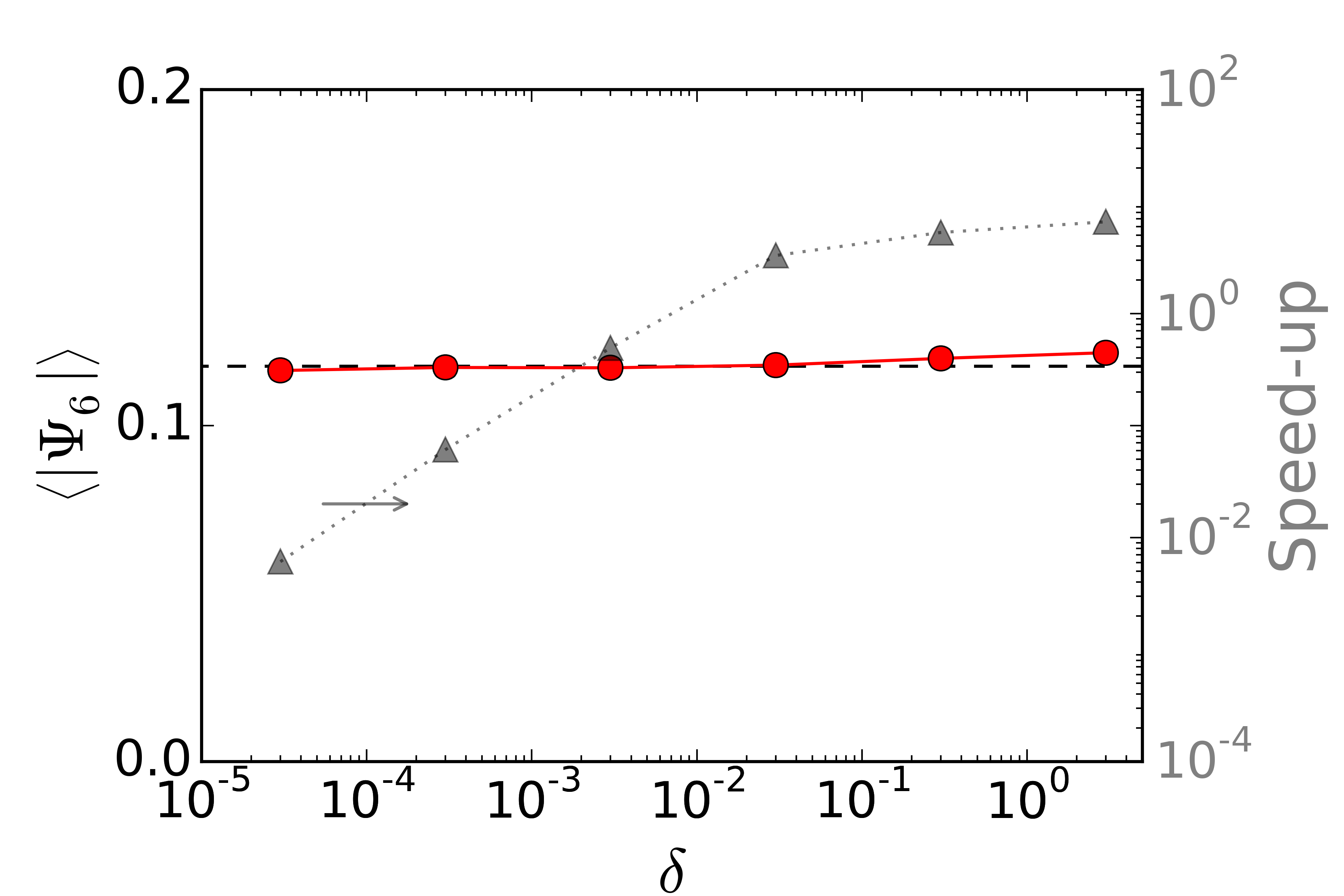}
\caption{\label{Psi_av} (color online). Results of the mean global orientational order parameter $\left< \vert\Psi_6\vert \right> $ by using the MCMD as a function of the step size $\delta$ (circles) and by using the event-driven MD (dashed line). On the secondary y-axis, the corresponding relative speed-up of the MCMD program vs. the MD program is shown (triangles). Parameters and initial conditions: $N$ = 64, density $\eta$ = 0.3, square lattice configuration with velocities sampled from uniform distribution and rescaled to $E_k/N = 10^{-4}$.}
\end{figure}

Before validating the dynamics, the method was verified in terms of static properties. First, the mean value of the global orientational order parameter

\begin{equation}\label{Equation10}
\Psi_6 = \frac{1}{N} \sum_{j=1}^N \sum_{l=1}^{N_{nb,j}} \frac {1} {N_{nb,j}}\exp(6i\phi_{jl})
\end{equation}

was considered. In Eq.~\ref{Equation10}, $N_{nb,j}$ is the number of neighbor disks for disk $j$ (calculated with a cut-off distance of 2.8 times the disk radius), $i$ is the imaginary unit and $\phi_{jl}$ is the angle with respect to the x-axis of the distance vector from the center of disk $j$ to the center of disk $l$. In Fig.~\ref{Psi_av} the mean value of $\vert \Psi_6 \vert $ calculated with MCMD is plotted as a function of the step size $\delta$ and compared with the corresponding MD result. The agreement between MCMD and MD is excellent in the whole range of $\delta$, spanning five orders of magnitude up to the largest value used, thus indicating that the method is robust.
At this point, some preliminary indications about the relative computational time required by MCMD and MD can be drawn. Even if one has to keep in mind that the aim of the proposed algorithm was not, in its current version, to optimize the computational efficiency, but rather to verify the possibility to calculate physically meaningful timescales in molecular systems. Hence the codes used focused more on readability than on computational efficiency. Nevertheless, it was found that for $\delta$ larger than approximately 0.01, MCMD achieves speed-up levels between 4 and 7 relative to event-driven MD, as shown in the secondary y-axis in Fig.~\ref{Psi_av}. This speed-up comes at the expense of some discrepancy between MCMD and MD, but the agreement remains acceptable (relative error 3.4\%) up to the largest value of displacement used ($\delta=3$), thus indicating that MCMD has a good potential from the view point of computational efficiency.

\begin{figure}
\includegraphics[scale=0.55]{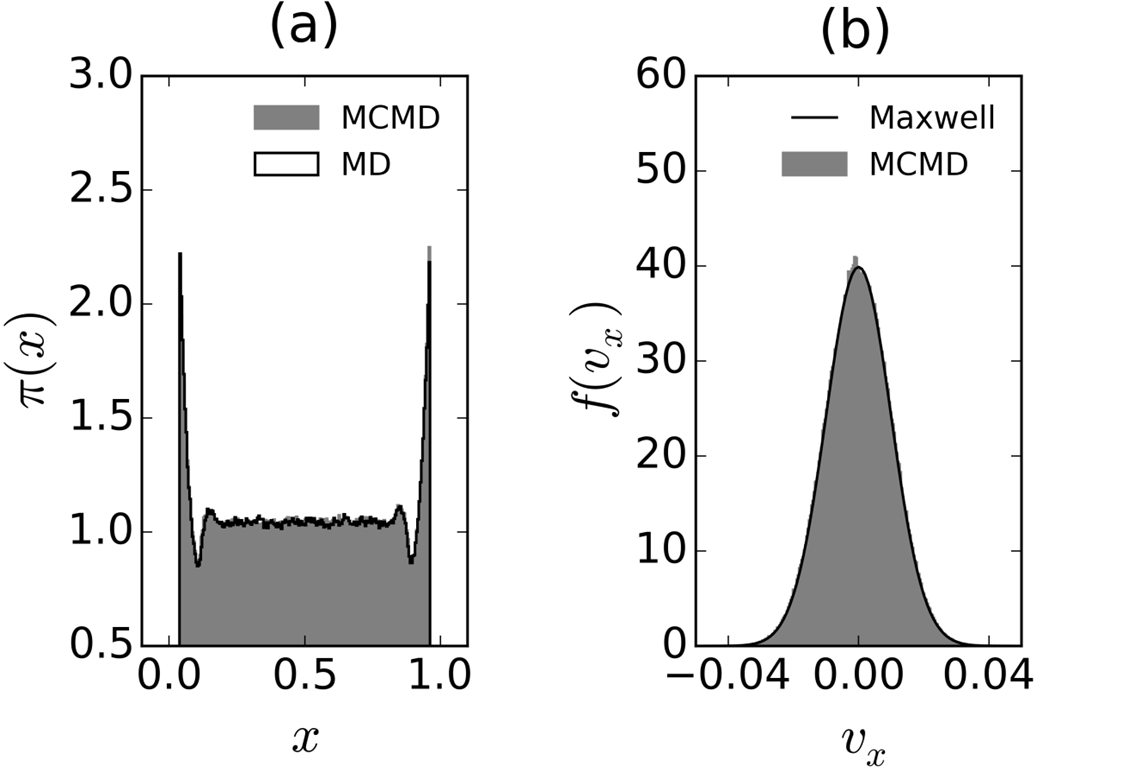}
\caption{\label{distributions} (color online). Equilibrium distributions calculated from a simulation run to $t$ = 35’000 with $\delta$ = 0.001 and $E_k/N = 10^{-4}$. (a) Normed histogram of the probability of having a disk at $x$, $\pi(x)$, by using the MCMD (filled gray histogram, 200 bins) and by using the event-driven MD (empty histogram, black line). (b) Comparison of the velocity distribution in the $x$ direction, $f(v_x)$, calculated with MCMD (filled gray histogram, 50 bins) and the Maxwell velocity distribution (black line).}
\end{figure}

The position probability $\pi(x)$, defined as the probability to find a disk at position $x$, is plotted in Fig.~\ref{distributions}a. The nonuniform $\pi(x)$ induced by the finite box size and by the entropic depletion interactions \cite{KrauthSMAC} is sampled correctly: the oscillations of $\pi(x)$ close to the wall boundaries match the MD results in terms of location and intensity of the maxima and minima. Remarkably, in Fig.~\ref{distributions}b, it is shown that the algorithm restores the equilibrium Maxwell velocity distribution \cite{Haile1997}. Obviously, the MD simulations provides exactly the same velocity distribution (not shown). It is worth stressing that the equilibrium distributions are obtained from the natural evolution of the system once the initial conditions are given, i.e., as the result of a dynamic trajectory through the phase space. As such, MCMD is very similar to MD: the equilibrium distributions of the microcanonical ensemble result from the ergodicity of the system rather than being obtained imposing the detailed balance at constant temperature, as in MMC. Another difference from traditional MMC is that the velocity distributions are obtained directly, again as in MD, without any need of independent samplings \cite{KrauthSMAC}.

\noindent \textbf{4.2 Dynamic Properties}

As an example of dynamic properties, the relaxation time, $\tau_R$, of the autocorrelation functions of the global order parameter and of the disk 1 velocity is considered. The two autocorrelation functions are defined as:
\begin{equation}\label{Equation11}
C_{\Psi_6}(t)= \frac {\left<\vert \Psi_6(t_0) \vert \vert \Psi_6(t_0+t) \vert \right> - \left<\vert \Psi_6 \vert\right>^2} {  \left< \vert \Psi_6 \vert ^2 \right> - \left< \vert \Psi_6 \vert \right> ^2}
\end{equation}

\begin{equation}\label{Equation12}
C_{v_1}(t)= \frac {\left< \textbf{\textit{v}}_1(t_0)\textbf{\textit{v}}_1(t_0+t) \right>} {\left< \textbf{\textit{v}}_1^2 \right>}
\end{equation}

\begin{figure}
\includegraphics[scale=0.55]{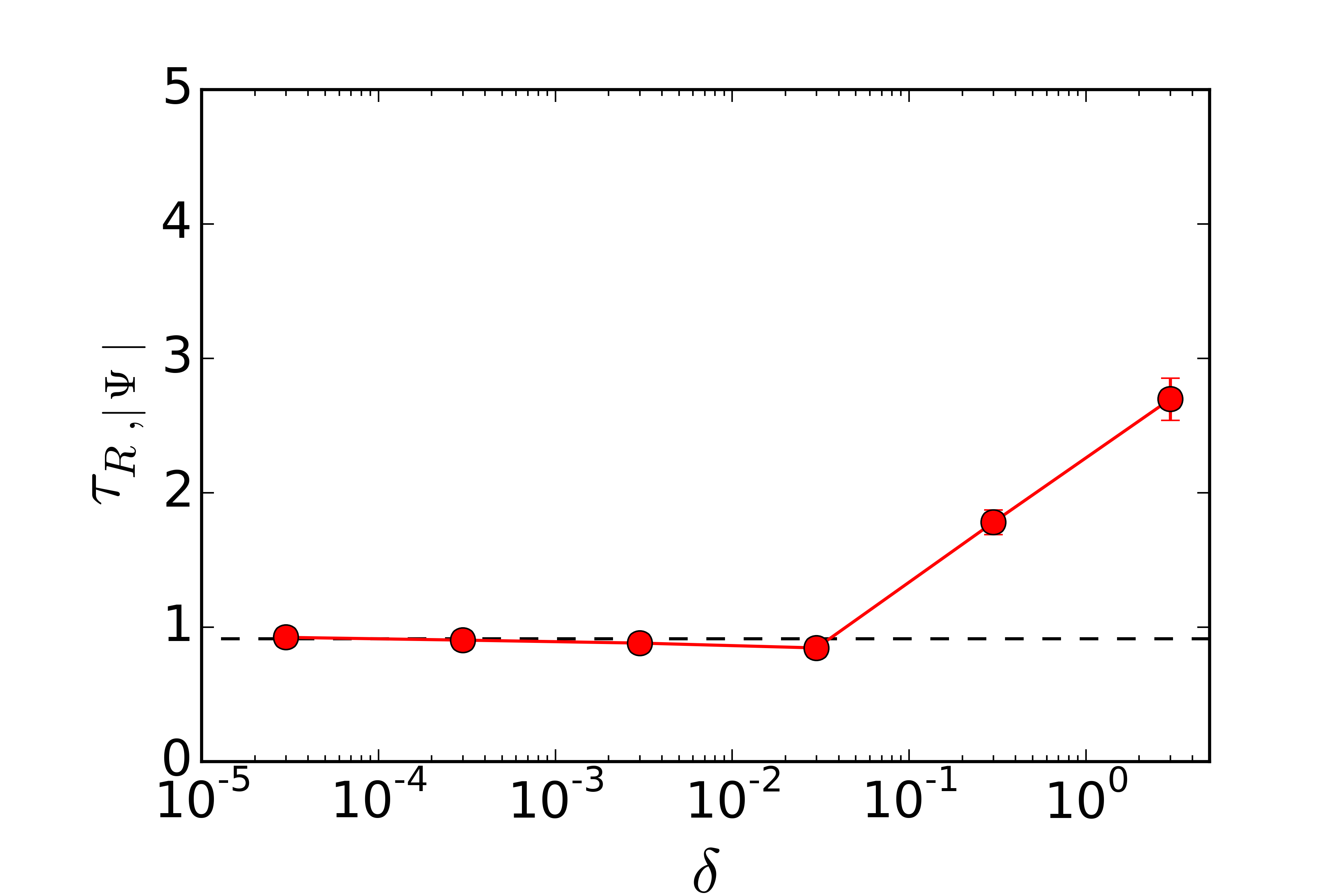}
\caption{\label{tau_Psi} (color online). Results of the relaxation time of mean global orientational order parameter, $\tau_{R,\vert\Psi \vert}$, by using the MCMD as a function of the step size $\delta$ (circles) and by using the event-driven MD (dashed line). The error bars represent the standard deviation of repeated simulation runs. Parameters and initial conditions: $N$ = 64, density $\eta$ = 0.3, square lattice configuration with velocities sampled from uniform distribution and rescaled to $E_k/N = 10^{-4}$.}
\end{figure}

\begin{figure}
\includegraphics[scale=0.55]{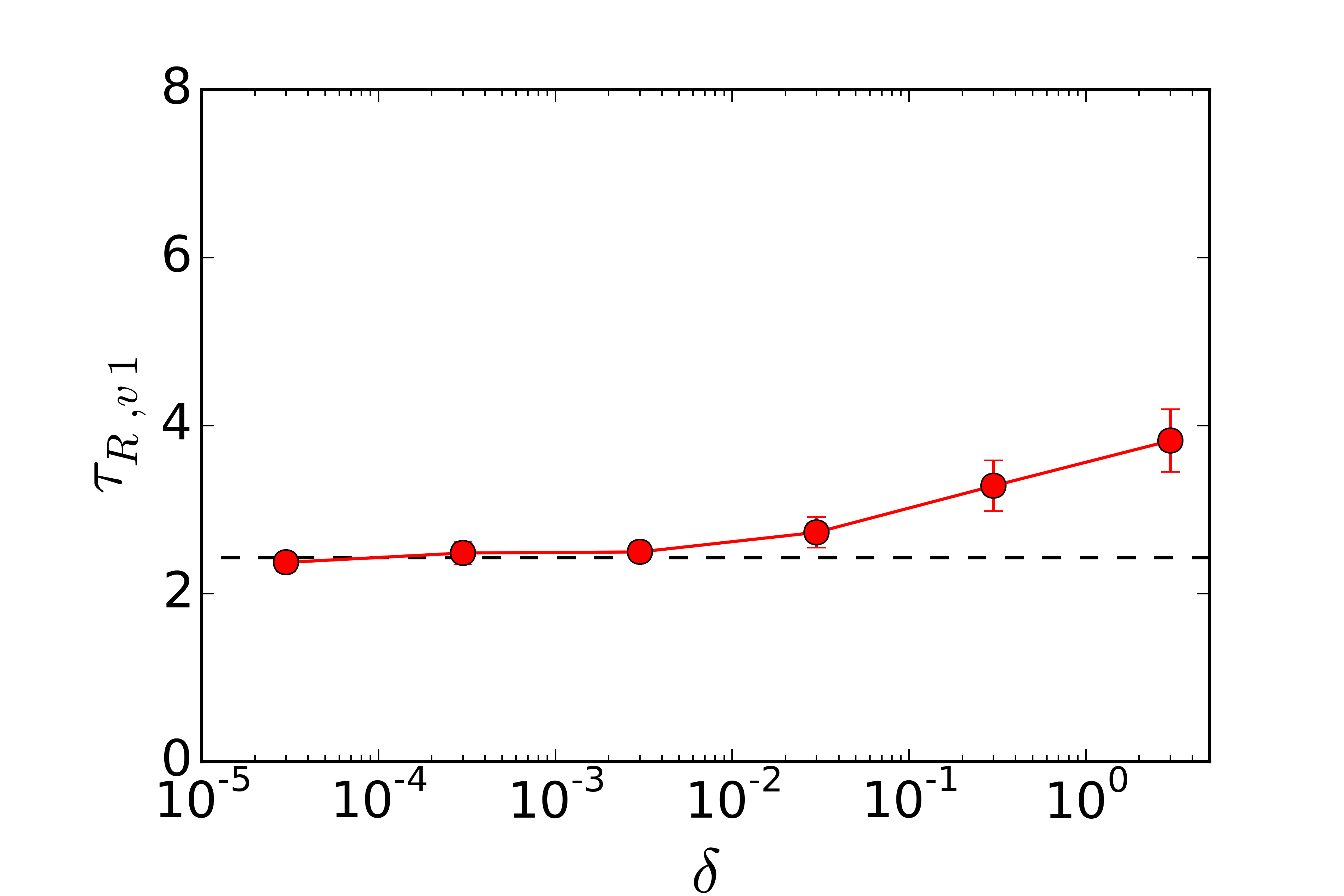}
\caption{\label{tau_v} (color online). Results of the relaxation time of the velocity of one disk, $\tau_{R,v1}$, by using the MCMD as a function of the step size $\delta$ (circles) and by using the event-driven MD (dashed line). The error bars represent the standard deviation of repeated simulation runs. Parameters and initial conditions: $N$ = 64, density $\eta$ = 0.3, square lattice configuration with velocities sampled from uniform distribution and rescaled to $E_k/N = 10^{-4}$.}
\end{figure}

The corresponding relaxation times, calculated as $C(\tau_R)$ = 1/$e$ \cite{Murat1996, Neelov1995}, are plotted in Fig.~\ref{tau_Psi} and Fig.~\ref{tau_v}, respectively, as a function of the step size. For decreasing displacement both $\tau_{R,\vert \Psi \vert}$ and $\tau_{R,v1}$ converge to the correct result, with an error relative to the MD rapidly decaying as the step size is decreased. The order of magnitude of both relaxation times was perfectly in line with the MD results even for the largest value of $\delta = 3$ tested. Under such conditions, $\delta$ was even larger than the box length (equal to 1), and therefore most of the disk displacement attempts led to collisions and MCMD provided disk trajectories significantly different from those obtained by MD. Nevertheless, the obtained results indicate that the method is fully reliable in providing meaningful timescales.

\begin{figure}
\includegraphics[scale=0.55]{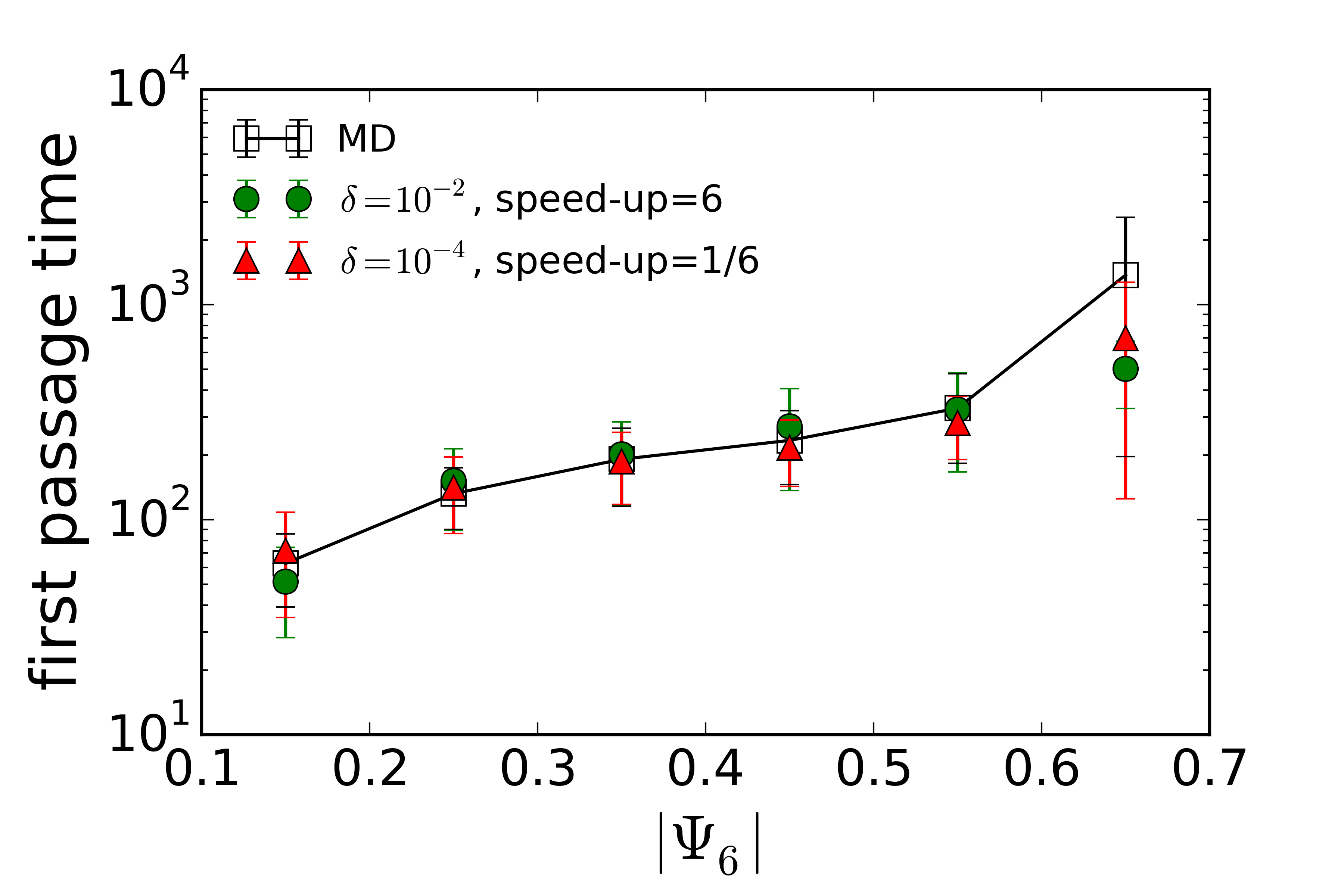}
\caption{\label{fpt} (color online). First passage time (y-axis) to a defined structural order (x-axis) for a system of  $N$ = 64 hard disks at $\eta$ = 0.72, starting from a square lattice configuration with velocities sampled from a Gaussian distribution and  $E_k/N = 10^{-5}$: MD (empty squares), MCMD with $\delta$ = 0.01 (circles), MCMD with $\delta$ = 0.0001 (triangles).}
\end{figure}

As a last example, a case of non-equilibrium dynamics at high density, $\eta$ = 0.72, is considered. Starting from a square lattice configuration, the order parameter increases from $\vert \Psi_6 \vert$ = 0 eventually fluctuating around the equilibrium value (cf. Supplementary Information \cite{Supplemental}). Fig.~\ref{fpt} shows the corresponding first passage times to configurations of increasing structural order, i.e., at which instant the system reaches, for the first time, a certain value of $\vert \Psi_6 \vert$. MCMD was run with values of $\delta$ that provided computational times comparable with MD, specifically with $\delta$ = 0.01 and $\delta$ = 0.0001, corresponding to approximately six times faster and six times slower than MD, respectively. In both cases the first passage times are in reasonable agreement with MD: considering the large standard deviation, one can state that the two methods give essentially the same results.\\

\noindent \textbf{5. Conclusions}

In summary, MCMD implements a rejection-free kinetic Monte Carlo scheme where the displacements are parallel to the velocities and the atoms are sampled proportionally to their velocities. With this choice, a physical timescale is naturally included in the method. Comparison with event-driven MD confirmed that MCMD gives direct access to correct dynamical quantities, including the velocity distributions, under both equilibrium and non-equilibrium conditions. The possibility of simulating the correct dynamics of molecular systems with atomistic detail broadens the scope of MC numerical tools. It is worth mentioning that by correct dynamics it is not meant that MCMD reproduces the very same trajectories for any atom as those that one would obtain from MD. Rather, it is meant that, when compared to MD, the method provides a reliable estimation of average dynamical quantities and timescales for the system as a whole, as shown in this contribution with the estimation of relaxation times and with the first passage time example.

Even if limited to the case of hard core interactions, the preliminary results reported in this letter are encouraging in view of future generalizations to soft potentials because they indicate that MCMD can also be a competitive alternative to molecular dynamics in terms of computational efficiency.
The most natural step forward is the introduction of still discrete but distance-dependent pair potentials, as in discrete molecular dynamics. In discrete molecular dynamics, continuous soft potentials (of Lennard-Jones type or even more complicated) are approximated by stepwise functions. Compared to traditional MD where the Newton's equations of motion must be solved, this approximation allows to solve the dynamics of the system using an event driven algorithm where atoms move with constant velocity until a step in the potential energy is crossed, after which the velocity of the atoms changes instantaneously \cite{Proctor2011}. This ballistic approach appears thus, at least in principle, fully compatible with MCMD. The main difference of an extended version of the MCMD compared to the current one would be represented by the fact that the Markov Chain process (steps 2 to 5a of the current algorithm) would proceed until a potential energy step is crossed, rather than until a collision between atoms or atoms and walls (step 5b) happens. Clearly, it would also be required to compute the distances between atoms and potential step changes accordingly. Considering that the use of discrete potentials enables to extend the timescale accessible to simulations \cite{Proctor2011}, a current central topic in the field of computational physics \cite{Salvalaglio2014, Henkelman2001, Tiwary2013, Voter1997, Proctor2011}, the possibility to generalize MCMD to the case of atoms interacting with distance dependent potentials approximated by discrete step functions is definitely of special interest.

Lastly, it can be noted that MCMD shares some similarities with the time-stamped force-bias MC (tfMC) \cite{Bal2014}, which also includes a timescale depending on the ratio between mean displacement and mean velocity similar to the one proposed here. tfMC shows great potential because of its ability to speed-up significantly the convergence towards equilibrium \cite{Bal2014, Neyts2012}. However, Bal and Neyts have recently performed a detailed analysis of the dynamics obtained by tfMC for various systems, including Lennard-Jones liquids, surface diffusion and defected graphene sheets, from which they concluded that the tfMC dynamics do not, in general, match the actual dynamics \cite{Bal2014}. One can reasonably speculate that the inability of the tfMC to converge to the correct dynamics is because particle displacements are biased in the direction of the force. In contrast, in MCMD, the displacements are forced in the direction of the velocities, a choice that mimics more closely the deterministic evolution of physical systems. Additionally, MCMD is extremely simple and maintains the same basic algorithmic structure of a conventional MMC scheme. This opens the way for the development of hybrid algorithms where equilibrium configurations are sampled with computationally efficient methods, e.g., tfMC, and the actual dynamics with MCMD only for those conditions that are of major interest.
With respect to computational efficiency, the algorithm proposed here compares favorably with event-driven MD. Additionally, the possibility to complement MCMD with a chain-event mechanism, thus reducing the number of disks samplings, is currently under investigation and might further speed-up the algorithm.

\textit{Acknowledgments.} Prof. W. Krauth and his research team at \'Ecole Normale Sup\'erieure, Paris, are gratefully acknowledged for having provided efficient code snippets.

\newpage
\noindent \textbf{Supplementary Information for :}\\

\noindent \textbf{"Meaningful timescales from Monte Carlo simulations of molecular systems with hard-core interactions"}\\

\noindent \textbf{Computational Details.} Simulations were performed with $N = 64$ disks of mass 1 at density $\eta = 0.3$ in a 2d square. Initial conditions for each simulation were a square lattice configuration with velocity components $v_{ix}$ and $v_{iy}$ sampled from a uniform distribution. Note that the selected initial conditions correspond, on purpose, to non-equilibrium conditions. Velocities were rescaled to have the same kinetic energy in all simulations, equal to $E_k/N = 10^{-4}$. The system was equilibrated for 500 time units, corresponding to more than 500 times the relaxation time of the order parameter, and data were collected every $\Delta t = 1$ for 3000 time units. The results plotted in Figure 1, Figure 3 and Figure 4 are averages of repeated simulation runs. The distributions plotted in Figure 2 are obtained from a single run to $t = 35’000$ with $\delta = 0.001$.
The first passage times plotted in Figure 5 are averages of repeated simulation runs with: $N = 64$, mass equal to 1, density $\eta = 0.72$ in a 2d square, square lattice as initial configuration (Figure S1) and initial velocity components $v_{ix}$ and $v_{iy}$ sampled from a gaussian distribution and rescaled to $E_k/N = 10^{-5}$.

\newpage
\noindent \textbf{Supplementary Scheme S1}\\

\begin{figure}[h]
\includegraphics[scale=1]{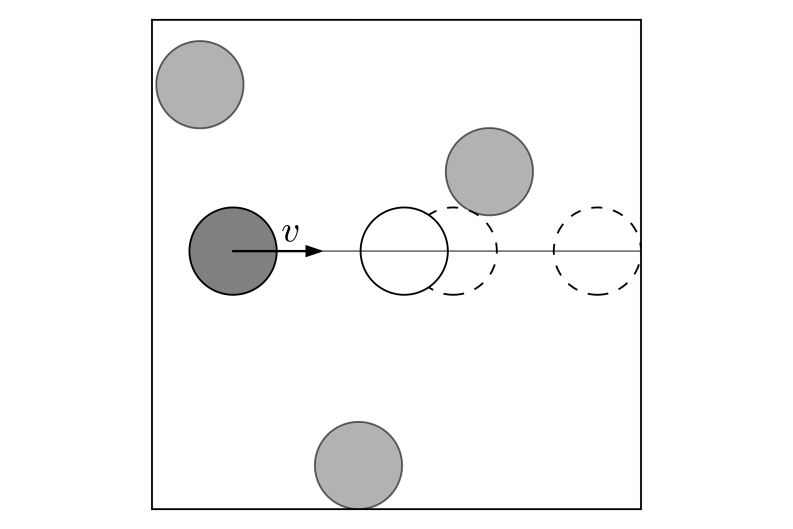}
\end{figure}
\noindent \textbf{Scheme S1.} Example of a Monte Carlo Molecular Dynamics simulation step with $N = 4$ hard
disks in a square box and displacement smaller than the minimum distance to a collision: $r_1\delta <$ min($\delta_{wall}, \delta_{pair})$. Given a configuration (grey disks), a disk $\mu$ is sampled (dark grey) and if the new position corresponding to the displacement $r_1\delta$ in the direction parallel to its velocity (white disk), is smaller than the minimum of the distances to the next pair and wall collisions (occurring in correspondence of the dashed disks), the disk is displaced to its new position. The minimum distances are calculated keeping all disks except the sampled one in their current position.

\newpage
\noindent \textbf{Supplementary Scheme S2}\\

\begin{figure}[h]
\includegraphics[scale=1]{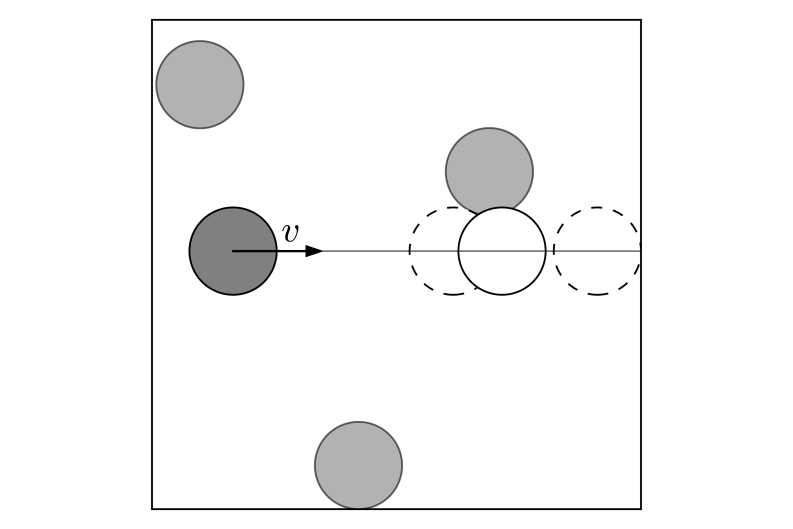}
\end{figure}
\noindent \textbf{Scheme S2.} Example of a Monte Carlo Molecular Dynamics simulation step with $N = 4$ hard disks in a square box and displacement larger than the minimum distance to a collision: $r_1\delta>$ min($\delta_{wall}, \delta_{pair})$. Given a configuration (grey disks), a disk $\mu$ is sampled (dark grey) and if the position corresponding to the displacement $r_1\delta$ in the direction parallel to its velocity (white disk), is larger than the minimum of the distances to the next pair and wall collisions (occurring in correspondence of the dashed disks), the disk is displaced of min($\delta_{wall}, \delta_{pair})$ only. The minimum distances are calculated keeping all disks except the sampled one in their current position. Velocities for the colliding disks are updated assuming perfectly elastic collisions.

\newpage
\noindent \textbf{Supplementary Figure S1}\\

\begin{figure}[h]
\includegraphics[scale=1]{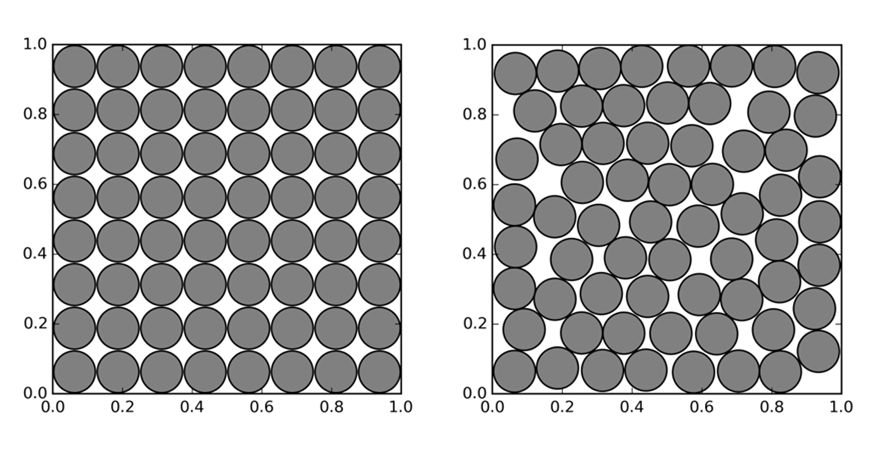}
\end{figure}
\noindent \textbf{Figure S1.} Initial (left) and final (right) configuration for a system of hard disks at $\eta$ = 0.72.

\newpage
\noindent \textbf{Supplementary Figure S2}\\

\begin{figure}[h]
\includegraphics[scale=0.8]{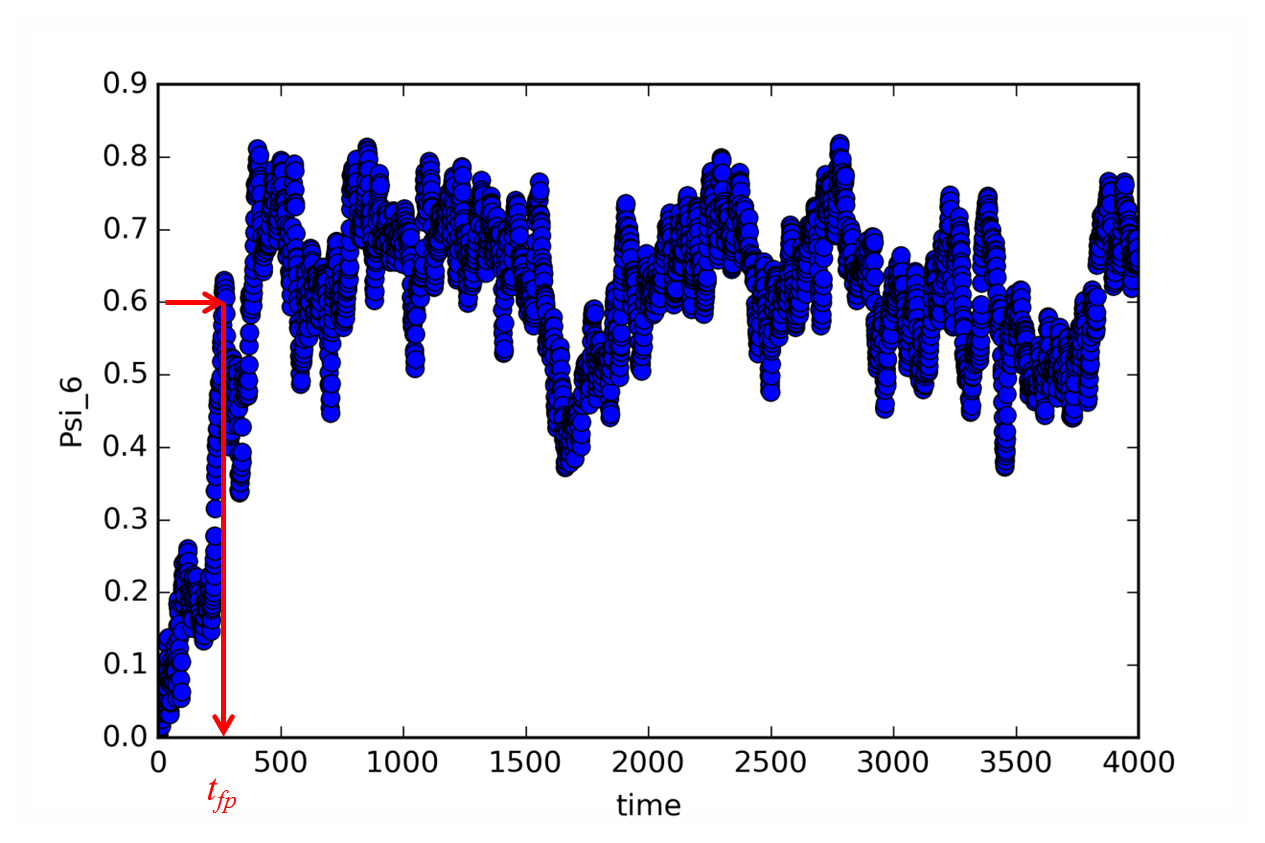}
\end{figure}
\noindent \textbf{Figure S2.} A dynamic trajectory of $\vert \Psi_6 \vert $ for $N = 64$ hard disks in a box at $\eta = 0.72$ and $E_k/N = 10^{-5}$. Initial configuration as in Figure S1. As an example, the red vertical arrow indicates the first passage time $t_{fp}$ to a structural ordering corresponding to $\vert \Psi_6 \vert $= 0.6.


\begin{thebibliography}{99}
\bibitem{note_tau} The SSA implements a Poisson process where the time step is sampled from an exponential distribution as $\tau$ = (1/$A$)ln(1/$r$), where $r$ is a random number drawn from the uniform distribution in the unit interval. The mean time step is $\left< \tau \right> $=1/$A$. Here, we are not interested in the temporal stochasticity and only the average time step is considered.
\end{thebibliography}
\end{document}